\documentstyle[epsf,aps,preprint]{revtex}
\tighten

\begin{document}
\draft

\bibliographystyle{prsty}

\title{Calculation of the Electron Self Energy for Low Nuclear Charge}

\author{Ulrich D.\ Jentschura,$^{1,2,}$\cite{InternetUJ}, 
Peter J.\ Mohr,$^{1,}$\cite{InternetPJM},
and Gerhard Soff$^{\,2,}$\cite{InternetGS}}

\address{$^1$National Institute of Standards and Technology, 
Gaithersburg, MD 20899-0001, USA \\
$^2$Institut f\"{u}r Theoretische Physik, TU Dresden,
Mommsenstra\ss e 13, 01062 Dresden, Germany}

\maketitle

\begin{abstract}
We present a nonperturbative numerical evaluation of the one-photon
electron self energy for hydrogenlike ions with low nuclear charge
numbers $Z=1$ to 5. Our calculation for the $1S$ state has a numerical
uncertainty of 0.8~Hz for hydrogen and 13~Hz for singly-ionized
helium. Resummation and convergence acceleration techniques that
reduce the computer time by about three orders of magnitude were
employed in the calculation.  The numerical results are compared to
results based on known terms in the expansion of the self energy in
powers of $Z\alpha$.
\end{abstract}


Recently, there has been a dramatic increase in the accuracy of
experiments that measure the transition frequencies in hydrogen and
deuterium \cite{BeEtAl1997,UdEtAl1997}.  This progress is due in part
to the use of frequency chains that bridge the range between optical
frequencies and the microwave cesium time standard.  The most
accurately measured transition is the $1S$-$2S$ frequency in hydrogen;
it has been measured with a relative uncertainty of $3.4 \times
10^{-13}$ or $840~{\rm Hz}$. With trapped hydrogen atoms, it should
be feasible to observe the $1S$-$2S$ frequency with an experimental
linewidth that approaches the $1.3\,{\rm Hz}$ natural width of the
$2S$ level \cite{CeEtAl1996,KiEtAl1998}.  Indeed, it is likely that
transitions in hydrogen will eventually be measured with an
uncertainty below $1\,{\rm Hz}$ \cite{HaPr,FrPr}.

In order for the anticipated improvement in experimental accuracy to
provide better values of the fundamental constants or better tests of
QED, there must be a corresponding improvement in the accuracy of the
theory of the energy levels in hydrogen and deuterium, particularly in
the radiative corrections that constitute the Lamb shift. As a step
toward a substantial improvement of the theory, we have carried out a
numerical calculation of the one-photon self energy of the $1S$ state
in a Coulomb field for values of the nuclear charge $Z = 1,2,3,4,5$.
This is the first complete calculation of the self energy at low $Z$
and provides a result that contributes an uncertainty of about 0.8~Hz
in hydrogen and deuterium. This is a decrease in uncertainty of more
than three orders of magnitude over previous results.

Among all radiative corrections, the largest by several orders of
magnitude are the one-photon self energy and vacuum polarization
corrections.  Of these, the larger and historically most problematic
is the self energy.  Analytic calculations of the electron self energy
at low nuclear charge $Z$ have extended over 50 years. The expansion
parameter in the analytic calculations is the strength of the external
binding field $Z\alpha$. This expansion is semi-analytic [i.e., it is
an expansion in powers of $Z\alpha$ and $\ln(Z\alpha)^{-2}$].  The
leading term was calculated in \cite{Be1947}. It is of the order of
$\alpha\,(Z\alpha)^4\,\ln(Z\alpha)^{-2}$ in units of $m_{\rm e}\,c^2$, 
where $m_{\rm e}$ is the mass of the electron. In subsequent work
\cite{Be1947,Fe1948,Fe1949,FrWe1949,KrLa1949,Sc1949,FuMiTo1949,Ba1951,
KaKlSc1952,BaBeFe1953,FrYe1958,FrYe1960,La1960,La1961ab,ErYe1965ab,
Er1971,Sa1981,Pa1993,Ka1997} higher-order coefficients were
evaluated.

The analytic results are relevant to low-$Z$ systems.  For high $Z$,
the complete one-photon self energy has been calculated without expansion
in $Z\alpha$ by numerical methods \cite{BrLaSa1959,DeJo1971,Mo1974a,
Mo1974b,Mo1975,ChJo1976,bas91,Mo1992,iam92,LiPeSaYn1993,MoSo1993,
ChJoSa1993}.
However, such numerical evaluations at low nuclear charge suffer from
severe loss of numerical significance at intermediate stages of the
calculation and slow convergence in the summation over angular
momenta.  As a consequence, the numerical calculations have been
confined to higher $Z$.

Despite these difficulties, the numerical calculations at higher $Z$
could be used together with the power-series results to extrapolate to
low $Z$ with an assumed functional form in order to improve the
accuracy of the self energy at low $Z$ \cite{Mo1975}; up to the
present, this approach has provided the most accurate theoretical
prediction for the one-photon self energy of the $1S$ state in
hydrogen \cite{Mo1996}.

However, this method is not completely satisfactory. The extrapolation
procedure gives a result with an uncertainty of 1.7~kHz, but employs a
necessarily incomplete analytic approximation to the higher-order
terms. It therefore contains a component of uncertainty that is
difficult to reliably assess. Termination of the power series at the
order of $\alpha\,(Z\alpha)^6$ leads to an error of 27~kHz. After the
inclusion of a result recently obtained in \cite{Ka1997} for the
logarithmic term of order $\alpha\,(Z\alpha)^7\,\ln(Z\alpha)^{-2}$ the
error is still 13~kHz.

A detailed comparison between the analytic and numerical approaches
has been inhibited by the lack of accurate numerical data for low
nuclear charge.  The one-photon problem is especially well suited for
such a comparison because five terms in the $Z\alpha$ expansion have
been checked in independent calculations. The known terms correspond
to the coefficients $A_{41}$, $A_{40}$, $A_{50}$, $A_{62}$ and
$A_{61}$ listed below in Eq.\ (\ref{coeffs}).

The energy shift $\Delta E_{\rm SE}$ due to the electron self energy is
given by
\begin{equation}
\label{defF}
\Delta E_{\rm SE} = \frac{\alpha}{\pi} \, 
\frac{(Z \alpha)^4}{n^3} \, m_{\rm e}c^2 \, F(Z\alpha)\,,
\end{equation}
where $n$ is the principal quantum number. For a particular atomic
state, the dimensionless function $F$ depends only on one argument,
the coupling $Z\alpha$. The semi-analytic expansion of $F(Z\alpha)$
about $Z\alpha=0$ gives rise to the following terms:
\begin{eqnarray}
\label{defGSE}
\lefteqn{F(Z\alpha) = A_{41} \, \ln(Z \alpha)^{-2} + A_{40} +
  (Z \alpha) \, A_{50} + (Z \alpha)^2} \nonumber \\
& & \;\; \times \left[A_{62} \, \ln^2(Z \alpha)^{-2} +
A_{61} \,\ln(Z \alpha)^{-2} + G_{\rm SE}(Z\alpha) \right]\,,
\end{eqnarray}
where $G_{\rm SE}(Z\alpha)$ represents the nonperturbative self-energy
remainder function.  The first index of the $A$ coefficients gives the
power of $Z\alpha$ [including the $(Z\alpha)^4$ prefactor from Eq.\
(\ref{defF})], the second corresponds to the power of the logarithm.
For the $1S$ ground state, which we investigate in this Letter, the
terms $A_{41}$ and $A_{40}$ were obtained in \cite{Be1947,Fe1948,
Fe1949,FrWe1949,KrLa1949,Sc1949,FuMiTo1949}. The correction term
$A_{50}$ was found in \cite{Ba1951,KaKlSc1952,BaBeFe1953}.  The
higher-order corrections $A_{62}$ and $A_{61}$ were evaluated and
confirmed in \cite{FrYe1958,FrYe1960,La1960,La1961ab,ErYe1965ab}.
The results are
\begin{eqnarray}
\label{coeffs}
A_{41} & = & \frac{4}{3}\,, \nonumber\\
A_{40} & = & \frac{10}{9} - \frac{4}{3}\,\ln k_0\,, \nonumber\\
A_{50} & = & 2\,\pi\,\left(\frac{139}{64} - \ln 2\right)\,,
\nonumber\\
A_{62} & = & -1\,, \nonumber\\
A_{61} & = & \frac{28}{3} \, \ln 2 - \frac{21}{20}\,. 
\end{eqnarray}
The Bethe logarithm $\ln k_0$ has been evaluated, e.g., in
\cite{KlMa1973,DrSw1990} as $\ln k_0 = 2.984~128~555~8(3)$.

For our high-accuracy, numerical calculation of $F(Z\alpha)$, we
divide the calculation into a high- and a low-energy part (see Ref.\
\cite{Mo1974a}).  Except for a further separation of the low-energy
part into an infrared part and a middle-energy part, which is
described in \cite{Je1998} and not discussed further here, we use the
same integration contour for the virtual photon energy and basic 
formulation as in \cite{Mo1974a}.

The numerical evaluation of the radial Green function of the bound
electron [see Eq.\ (A.16) in \cite{Mo1974a}] requires the calculation
of the Whittaker function $W_{\kappa,\mu}(x)$ (see \cite{MaObSo1966},
p.\ 296) over a very wide range of parameters $\kappa$, $\mu$ and
arguments $x$. Because of numerical cancellations in subsequent steps
of the calculation, the function $W$ has to be evaluated to 1 part in
$10^{24}$.  In a problematic intermediate region, which is given
approximately by the range $15 < x < 250$, we found that resummation
techniques applied to the divergent asymptotic series of the function
$W$ provide a numerically stable and efficient evaluation
scheme. These techniques follow ideas outlined in \cite{We1996c} and
are described in detail in \cite{Je1998}.

For the acceleration of the slowly convergent angular momentum sum in
the high-energy part [see Eq.\ (4.3) in \cite{Mo1974b}], we use the
combined nonlinear-condensation transformation \cite{JeMoSoWe1998}.
This transformation consists of two steps: First, we apply the van
Wijngaarden condensation transformation \cite{vW1965} to the original
series to transform the slowly convergent monotone input series into
an alternating series \cite{Da1969}.  In the second step, the
convergence of the alternating series is accelerated by the $\delta$
transformation [see Eq.\ (3.14) in \cite{JeMoSoWe1998}]. The $\delta$
transformation acts on the alternating series much more effectively
than on the original input series. The highest angular momentum,
characterized by the Dirac quantum number $\kappa$, included in the
present calculation is about $3~500~000$. However, even in these
extreme cases, evaluation of less than $1~000$ terms of the original
series is required.  As a result, the computer time for the evaluation
of the slowly convergent angular momentum expansion is reduced by
roughly three orders of magnitude.  The convergence acceleration
techniques remove the principal numerical difficulties associated with
the singularity of the relativistic propagators for nearly equal
radial arguments.  These singularities are present in all QED effects
in bound systems, irrespective of the number of photons involved. It
is expected that these techniques could lead to a similar decrease in
computer time in the calculation of QED corrections involving more
than one photon.

\begin{figure}[htb]
\caption{\label{fig1} The self-energy function $F(Z\alpha)$. The
points are the numerical results of this work, the curve is given by
the analytically known terms that correspond to the coefficients
listed in Eq.~(\ref{coeffs}).}
\centerline{\mbox{\epsfysize=6.0cm\epsffile{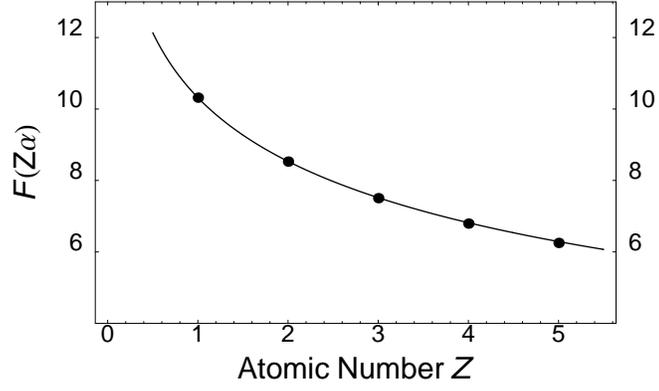}}}
\end{figure}

In the present calculation, numerical results are obtained for the
scaled self-energy function $F(Z\alpha)$ for the nuclear charges
$Z=1,2,3,4,5$ (see Table 1). The value of $\alpha$ used in the
calculation is $\alpha_0 = 1/137.036$. This is close to the current
value from the anomalous magnetic moment of the electron
\cite{Ki1998},
\[
1/\alpha = 137.035~999~58(52)\,. 
\]
The numerical data points are plotted in Fig.~\ref{fig1}, together
with a graph of the function determined by the analytically known
lower-order coefficients listed in Eq.~(\ref{coeffs}).

In order to allow for a variation of the fine-structure constant, we
repeated the calculation with two more values of $\alpha$, which are
\[
1/\alpha_> = 137.035~999~5 \;\;\; \mbox{and} \;\;\; 
1/\alpha_< = 137.036~000~5\,.
\]
On the assumption that the main dependence of $F$ on $Z\alpha$ is
represented by the lower-order terms in (\ref{coeffs}), the change in
$F(Z\alpha)$ due to the variation in $\alpha$ is
\[
\frac{\partial F(Z\alpha)}{\partial \alpha} \, \delta\alpha =
-2\,A_{41}\,\frac{\delta\alpha}{\alpha} +
\left[Z\,A_{50} + {\rm O}(\alpha\ln^2\alpha)\right] \,\delta\alpha
\]
for a given nuclear charge $Z$.  Based on this analytic estimate, we
expect a variation
\[
F(Z\alpha_>) - F(Z\alpha_0) \approx
F(Z\alpha_0) - F(Z\alpha_<) \approx
-9\times10^{-9}
\]
for the different values of $\alpha$.  This variation is in fact
observed in our calculation. E.g., for the case $Z=2$ we find
\begin{eqnarray}
F(2\alpha_{<}) &=& 8.528~325~061(1)\,, \nonumber\\
F(2\alpha_0)   &=& 8.528~325~052(1) \; \mbox{and}\nonumber\\
F(2\alpha_{>}) &=& 8.528~325~043(1)\,. \nonumber
\end{eqnarray}
This constitutes an important stability check on the numerics and it
confirms that the main dependence of $F$ on its argument is indeed
given by the lowest-order analytic coefficients $A_{41}$ and $A_{50}$.

In addition to the results for $F(Z\alpha_0)$, numerical results for
the nonperturbative self-energy remainder function $G_{\rm
SE}(Z\alpha_0)$ are also given in Table 1.  The results for the
remainder function are obtained from the numerical data for
$F(Z\alpha_0)$ by direct subtraction of the analytically known terms
corresponding to the coefficients $A_{41}$, $A_{40}$, $A_{50}$,
$A_{62}$ and $A_{61}$ [see Eqs.~(\ref{defGSE},\ref{coeffs})]. Note
that because the dependence of $F$ on $Z\alpha$ is dominated by the
subtracted lower-order terms, we have at the current level of accuracy
$G_{\rm SE}(Z\alpha_{<}) = G_{\rm SE}(Z\alpha_0) = G_{\rm
SE}(Z\alpha_{>})$. The numerical uncertainty of our calculaton is $0.8
\times Z^4 \, {\rm Hz}$ in frequency units.

A sensitive comparison of numerical and analytic approaches to the
self energy can be made by extrapolating the nonperturbative
self-energy remainder function $G_{\rm SE}(Z\alpha)$ to the point
$Z\alpha=0$.  It is expected that the function $G_{\rm SE}(Z\alpha)$
approaches a constant in the limit $Z\alpha\to 0$. This constant is
referred to as $G_{\rm SE}(0) \equiv A_{60}$.  In the analytic
approach, much attention has been devoted to the coefficient $A_{60}$
\cite{ErYe1965ab,Er1971,Sa1981,Pa1993}. The correction has
proven to be difficult to evaluate, and analytic work on $A_{60}$ has
extended over three decades.  A step-by-step comparison of the
analytic calculations has not been feasible, because the approaches to
the problem have differed widely. An additional difficulty is
the isolation of terms which contribute in a given order in $Z\alpha$,
i.e. the isolation of only those terms which contribute to $A_{60}$
(and not to any higher-order coefficients).

In order to address the question of the consistency of $A_{60}$ with
our numerical results, we perform an extrapolation of our data to the
point $Z\alpha=0$.  The extrapolation procedure is adapted to the
problem at hand. We fit $G_{\rm SE}$ to an assumed functional form
which corresponds to $A_{60}$, $A_{71}$ and $A_{70}$ terms, with the
coefficients to be determined by the fit. We find that our numerical
data is consistent with the calculated value $A_{60} = -30.924\,15(1)$
\cite{Pa1993,PaPr}. It is difficult to assess the seventh-order
logarithmic term $A_{71}$, because the extrapolated value for $A_{71}$
is very sensitive to possible eighth-order triple and double
logarithmic terms, which are unknown. We obtain as an approximate
result $A_{71} = 5.5(1.0)$, and we therefore cannot conclusively confirm
the result \cite{Ka1997}
\[
A_{71} = \pi\,\left(\frac{139}{64}-\ln2\right) = 4.65.
\]
Since our all-order numerical evaluation eliminates the uncertainty
due to higher-order terms, we do not pursue this question any
further.

\begin{figure}[htb]
\caption{\label{fig2} Results for the scaled self-energy
remainder function $G_{\rm SE}(Z\alpha)$ at low $Z$.}
\centerline{\mbox{\epsfysize=6.0cm\epsffile{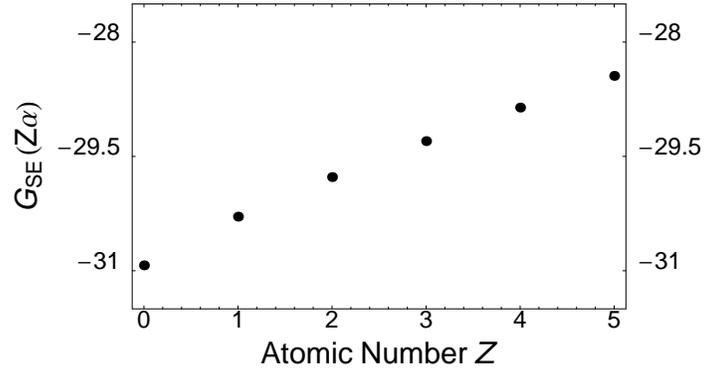}}}
\end{figure}

The numerical data points of the function $G_{\rm SE}(Z\alpha)$ are
plotted in Fig.\ \ref{fig2} together with the value $G_{\rm SE}(0) =
A_{60} = -30.924\,15(1)$.  For a determination of the Lamb shift, the
dependence of $G_{\rm SE}$ on the reduced mass $m_{\rm r}$ of the
system has to be restored. In general, the coefficients in the
analytic expansion (\ref{defGSE}) acquire a factor $(m_{\rm r}/m_{\rm
e})^3$, because of the scaling of the wave function.  Terms associated
with the anomalous magnetic moment are proportional to $(m_{\rm
r}/m_{\rm e})^2$ \cite{SaYe1990}. The nonperturbative remainder
function $G_{\rm SE}$ is assumed to be approximately proportional to
$(m_{\rm r}/m_{\rm e})^3$, but this has not been proved
rigorously. Work is currently in progress to address this question
\cite{PaKaPr}.

We conclude with a brief summary of the results of this Letter. (i) We
have obtained accurate numerical results for the self energy at low
nuclear charge. Previously, severe numerical cancellations have been a
problem for these evaluations.  (ii) For a particular example, we have
addressed the question of how well semi-analytic expansions represent
all-order results at low nuclear charge.  Our numerical data is
consistent with the value $A_{60}= -30.924\,15(1)$ \cite{Pa1993,PaPr}.
(iii) Numerical techniques \cite{JeMoSoWe1998} have been developed
that reduce the computer time for the problem by about three orders of
magnitude.

The calculation presented here is of importance for the interpretation
of measurements in hydrogen, deuterium and singly-ionized helium and
for the improvement of the Rydberg constant, because of recent and
projected progress in accuracy. In the determination of the Rydberg
constant, uncertainty due to the experimentally determined proton
radius can be eliminated by comparing the frequencies of more than one
transition \cite{UdEtAl1997}. We have shown that an all-order
calculation can provide the required accuracy if suitable numerical
methods are used.

The authors acknowledge helpful discussions with E.J.\ Weniger.
U.D.J. gratefully acknowledges helpful conversations with J.\ Baker,
J.\ Conlon, J.\ Devaney and J.\ Sims, and support by the Deutsche
Forschungsgemeinschaft (contract no.\ SO333/1-2) and the Deutscher
Akademischer Austauschdienst. P.J.M.\ is grateful to Rebecca Ghent who
participated in earlier exploratory work on this calculation, and he
acknowledges continued support by the Alexander-von-Humboldt
Foundation. G.S.\ acknowledges continued support by the Gesellschaft
f\"{u}r Schwerionenforschung and the Deutsche Forschungsgemeinschaft.

%
%

%
%

\begin{table}[htb]
\caption{\label{restable} Scaled self-energy function and
nonperturbartive self-energy remainder function
for low-$Z$ hydrogenlike systems.}
\begin{center}
\begin{minipage}{3.5in}
\begin{tabular}{lr@{.}lr@{.}l}
\multicolumn{5}{c}{\rule[-3mm]{0mm}{8mm}{$F(Z\alpha_0)$ and
$G_{\rm SE}(Z\alpha_0)$}} \\
$Z$ &
\multicolumn{2}{c}{$F(Z\alpha_0)$} &
\multicolumn{2}{c}{$G_{\rm SE}(Z\alpha_0)$} \\
\hline
$1$ \hspace{0.2in} &
  $10$ & $316~793~650(1)$ \hspace{0.2in} &
 $-30$ & $290~24(2)$ \\
$2$ &
   $8$ & $528~325~052(1)$ &
 $-29$ & $770~967(5)$ \\
$3$ &
   $7$ & $504~503~422(1)$ &
 $-29$ & $299~170(2)$ \\
$4$ &
   $6$ & $792~824~081(1)$ &
 $-28$ & $859~222(1)$ \\
$5$ &
   $6$ & $251~627~078(1)$ &
 $-28$ & $443~472~3(8)$ \\
\end{tabular}
\end{minipage}
\end{center}
\end{table}

%
%

%
%

%
%

\end{document}